\documentclass[twocolumn,aps,prl,floatfix]{revtex4}
\usepackage{amsmath,amssymb}
\usepackage[dvips]{graphicx}

\begin{document}

\title{Two-dimensional electrons at a cleaved semiconductor surface: \\
Observation of the quantum Hall effect}

\author{Yukihide Tsuji}
\author{Toshimitsu Mochizuki}
\author{Tohru Okamoto${}^{\rm {a)}}$\footnotetext{${}^{\rm {a)}}$Electronic mail: okamoto@phys.s.u-tokyo.ac.jp}}

\affiliation{Department of Physics, University of Tokyo, Hongo, Bunkyo-ku, Tokyo 113-0033, Japan}

\begin{abstract}
Low-temperature in-plane magnetotransport measurements have been performed on adsorbate-induced electron systems formed at {\itshape in-situ} cleaved surfaces of {\itshape p}-type InAs. The Ag-coverage dependence of the surface electron density strongly supports a simple model based on a surface donor level lying above the conduction band minimum. The observations of the quantized Hall resistance and zero longitudinal resistivity demonstrate the perfect two-dimensionality of the surface electron system. We also observed the Rashba effect due to the strong asymmetry of the confining potential well.

\end{abstract}

\maketitle

Two-dimensional electron systems (2DESs) exhibit various important transport phenomena, such as the integer and fractional quantum Hall effect,  \cite{Klitzing,Tsui} zero-magnetic field metal-insulator transition, \cite{Abrahams} and zero-resistance states induced by microwave radiation. \cite{Mani,Zudov} So far these phenomena have been studied for 2DESs formed within solid-state devices, where electrons are confined to semiconductor interfaces. \cite{Ando} At the surface of InAs, on the other hand, conduction electrons can be induced by submonolayer deposition of other materials. Photoelectron spectroscopy measurements have shown that the position of the Fermi level lies above the conduction band minimum at the surfaces of InAs with various adsorbates. \cite{Baier86,Aristov91,Aristov93,Aristov94,Morgenstern00,Getzlaff} Research on electrons at semiconductor surfaces has great future potential because of the variety of adsorbates and the application of scanning probe microscopy techniques. \cite{Morgenstern03} Most of the previous photoelectron spectroscopy and scanning tunneling spectroscopy measurements have been done on $n$-type InAs crystals where electrical contact to 2D-like surface electrons was made through the substrate. In order to investigate 2D transport properties, however, the surface electrons should be separated from 3D carriers in the substrate. In the case of $p$-type crystals, the isolation is expected to be achieved by the depletion layer. In this work, we have successfully performed the first in-plane magnetotransport measurements on {\it in-situ} cleaved surfaces of $p$-type InAs. The Hall measurements enabled us to determine the surface electron density as a function of the coverage of Ag. The perfect two-dimensionality of the adsorbate-induced surface electron system was established by the observation of the quantum Hall effect.

The samples used were cut from a Zn-doped single crystal with an acceptor concentration of $1.3 \times 10^{17}~{\rm cm}^{-3}$. Sample geometry is shown in Fig. 1. Two current electrodes and four voltage electrodes were prepared at room temperature by deposition of gold films onto non-cleaved surfaces. Sample cleaving, subsequent deposition of Ag atoms and magnetotransport measurements on the cleaved (110) surface were performed at low temperatures in an ultra-high vacuum chamber with a ${}^4$He cryostat. A magnetic field was applied in the perpendicular direction with respect to the cleaved surface and the conduction electron density $N_s$ can be accurately determined from the Hall coefficient. The width of the Hall bar (wafer thickness) is 0.4~mm and the longitudinal resistivity $\rho_{xx}$ was measured from the voltage difference between two electrodes separated by 0.8~mm. Before cleaving, leakage resistance between any two electrodes is greater than or on the order of $10^6~{\rm \Omega}$ at 1.5~K, while it often decreases to the order of $10^5~{\rm \Omega}$ at 4.2~K. The magnetotransport measurements were made at our lowest temperature of 1.5~K to eliminate the parallel conduction through the non-cleaved surfaces or the $p$-type substrate.
All the data shown here were taken on optically flat surfaces.
\begin{figure}
\includegraphics[width=8cm]{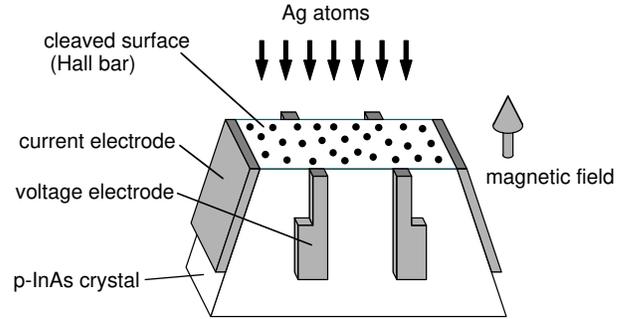}
\caption{
Hall bar geometry for magnetotransport measurements on a (110) cleaved surface. Six gold electrodes were prepared on non-cleaved surfaces in advance. Sample cleaving, deposition of Ag atoms, and magnetotransport measurements were performed under low temperature and ultra-high vacuum conditions.
}
\end{figure}

The conductance of the cleaved surface is very low before the deposition of Ag atoms. The in-plane conduction is obtained after a tiny deposition. Figure 2 shows the dependence of $N_s$ on the Ag-coverage $\Theta$ in the low $\Theta$ region. One monolayer (ML) is defined as Ag atomic density equivalent to the surface atomic density of InAs(110) ($7.75 \times 10^{14}~{\rm atoms} / {\rm cm}^2$), and the values of $\Theta$ in this work have uncertainties of 10~\%. $N_s$ increases almost linearly with $\Theta$ until it reaches the maximum value of $\approx 3.3 \times 10^{12}~{\rm cm}^{-2}$. The solid line is calculated taking into account the depletion layer charge \cite{Ando} on the assumption that each adsorbed Ag atom donates one electron to InAs. It reproduces the observed initial linear increase. The saturation of $N_s$ can be attributed to a pinning of the surface Fermi level $\varepsilon_{\rm F}$ at a Ag-induced surface donor level \cite{Monch} located in the conduction band. The position of $\varepsilon_{\rm F}$ above the conduction-band minimum $\varepsilon_{\rm C}$ is calculated taking into account the nonparabolicity of the InAs conduction band \cite{Merkt} (see the right axis of Fig. 2). The maximum value of $\varepsilon_{\rm F}-\varepsilon_{\rm C} \approx 0.38~{\rm eV}$ agrees with a peak of the $\varepsilon_{\rm F}$ position obtained from photoelectron spectroscopy measurements on a Ag-deposited (110) surface of $p$-InAs at 20~K. \cite{Aristov93}
\begin{figure}
\includegraphics[width=8cm]{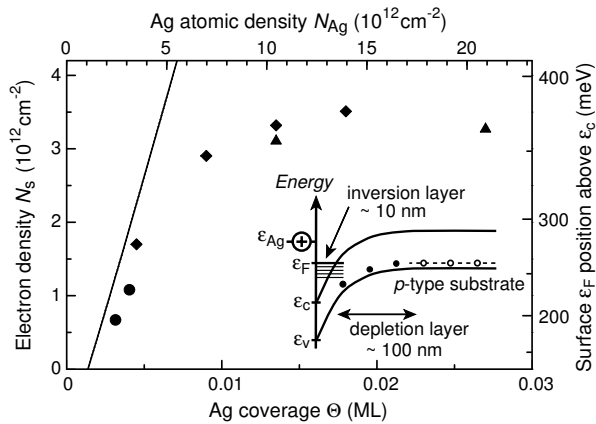}
\caption{
Electron density $N_s$ in the low Ag-coverage region. Different symbols refer to different samples. The solid line represents a calculation for $N_s = N_{\text {Ag}} - N_{\text {depl}}$, where $N_{\text {Ag}}$ is the Ag atomic density and $N_{\text {depl}}$ is the depletion layer charge per unit area ($\sim 10^{12}~{\text {cm}}^{-2}$). The Fermi energy position $\varepsilon_{\text F}$ above the conduction band minimum $\varepsilon_{\text C}$ at the surface (right axis) is estimated as a function of $N_s$ using the approximation derived in Ref.~\onlinecite{Merkt}. Inset: Schematic energy diagram of the surface. $\varepsilon_{\text V}$ is the valence band maximum and the band gap $\varepsilon_{\text C} - \varepsilon_{\text V}$ of InAs is 417~meV. Open and filled circles represent neutral and ionized acceptors, respectively. When the 2D electrons in the inversion layer are filled up to Ag-induced surface donor level $\varepsilon_{\text {Ag}}$, $\varepsilon_{\text F}$ reaches the maximum.
}
\end{figure}

Figure 3 shows the overall $\Theta$-dependence of $N_s$ and the electron mobility $\mu$. $N_s$ shows negative $\Theta$-dependence for $\Theta \gtrsim 0.04$. The decrease in $N_s$ may correspond to the drop of $\varepsilon_{\rm F}$ observed in the photoelectron spectroscopy measurements. \cite{Aristov91,Aristov93} The formation of the clusters of Ag might reduce the surface donor level below that for isolated adsorbates. \cite{Aristov91,Aristov93} The electron mobility $\mu$ increases together with $N_s$ for $\Theta \lesssim 0.01$. This can be understood in terms of negative $N_s$-dependence of the scattering rate due to screened Coulomb potentials. \cite{Stern} In the higher $\Theta$ region, on the other hand, $\mu$ does not show a clear dependence on $\Theta$ and the relationship between $\mu$ and $N_s$ is different from that observed in the low $\Theta$ region. The results indicate that the scattering rate has a direct dependence on $\Theta$ and the ionized adsorbates act as scattering centers at least in the low $\Theta$ region. The potential fluctuations due to the adsorbates are expected to decrease with increasing $\Theta$ because of the reduction of the average ionization probability of each adsorbed atom. \cite{Getzlaff}
\begin{figure}
\includegraphics[width=8cm]{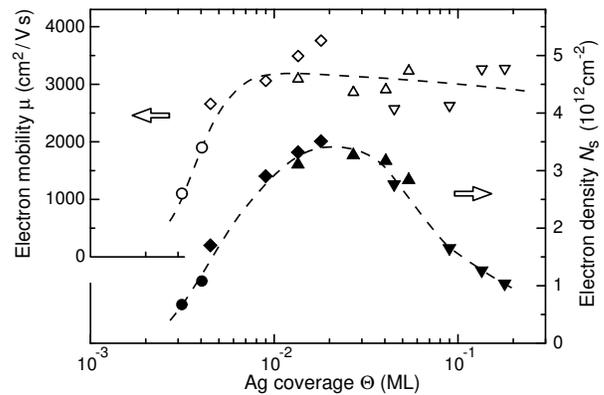}
\caption{
Electron density (filled symbols, right axis) and electron mobility (open symbols, left axis) as functions of Ag coverage. Dotted curves are guides to the eye.
}
\end{figure}

The two-dimensionality of the Ag-induced electron system was established by the observation of the quantum Hall effect. In high magnetic fields, the Landau level separation $\hbar \omega_c$ exceeds the level broadening $\hbar \tau^{-1}$. Here $\omega_c$ is the cyclotron frequency, $\tau$ is the scattering time and the dimensionless parameter $\omega_c \tau$ is equivalent to $\mu B$ (1~T${}^{-1} = 10^{4}~{\rm cm}^2/{\rm Vs}$). In Fig.~4(a), $\rho_{xx} (B)$ and the Hall resistance $R_{\rm H} (B)$ for $\Theta=0.18~{\rm ML}$ with $N_s =1.04 \times 10^{12}~{\rm cm}^{-2}$ and $\mu=0.327~{\rm T}^{-1}$ are shown. SdH oscillations appear above 3~T for which $\mu B$ exceeds unity. $\rho_{xx}$ minima are observed for even-integer values of the Landau level filling factor $\nu=h N_s / eB$ as expected for spin degenerate 2DESs. The quantum Hall plateau for $\nu=4$ and the vanishing of $\rho_{xx}$ are observed around $B=10.8$~T. In contrast to 2DESs in solid-state devices, scanning probe microscopy techniques can be applied directly to the 2DES formed at the surface. \cite{Morgenstern03} Further investigations may reveal current distributions in a quantum Hall bar and the structure of edge channels. \cite{Chklovskii}

\begin{figure}
\includegraphics[width=8cm]{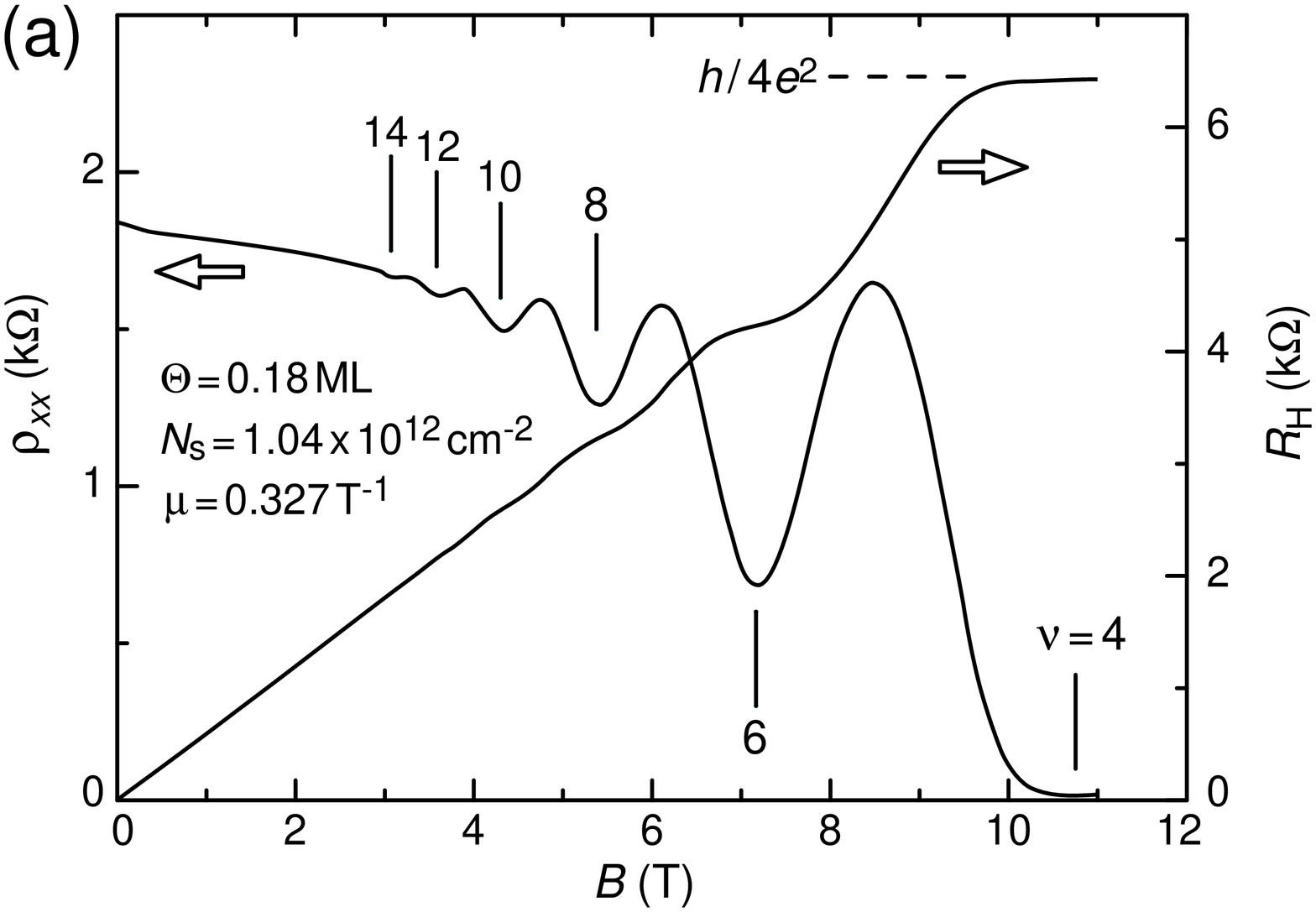}
\includegraphics[width=8cm]{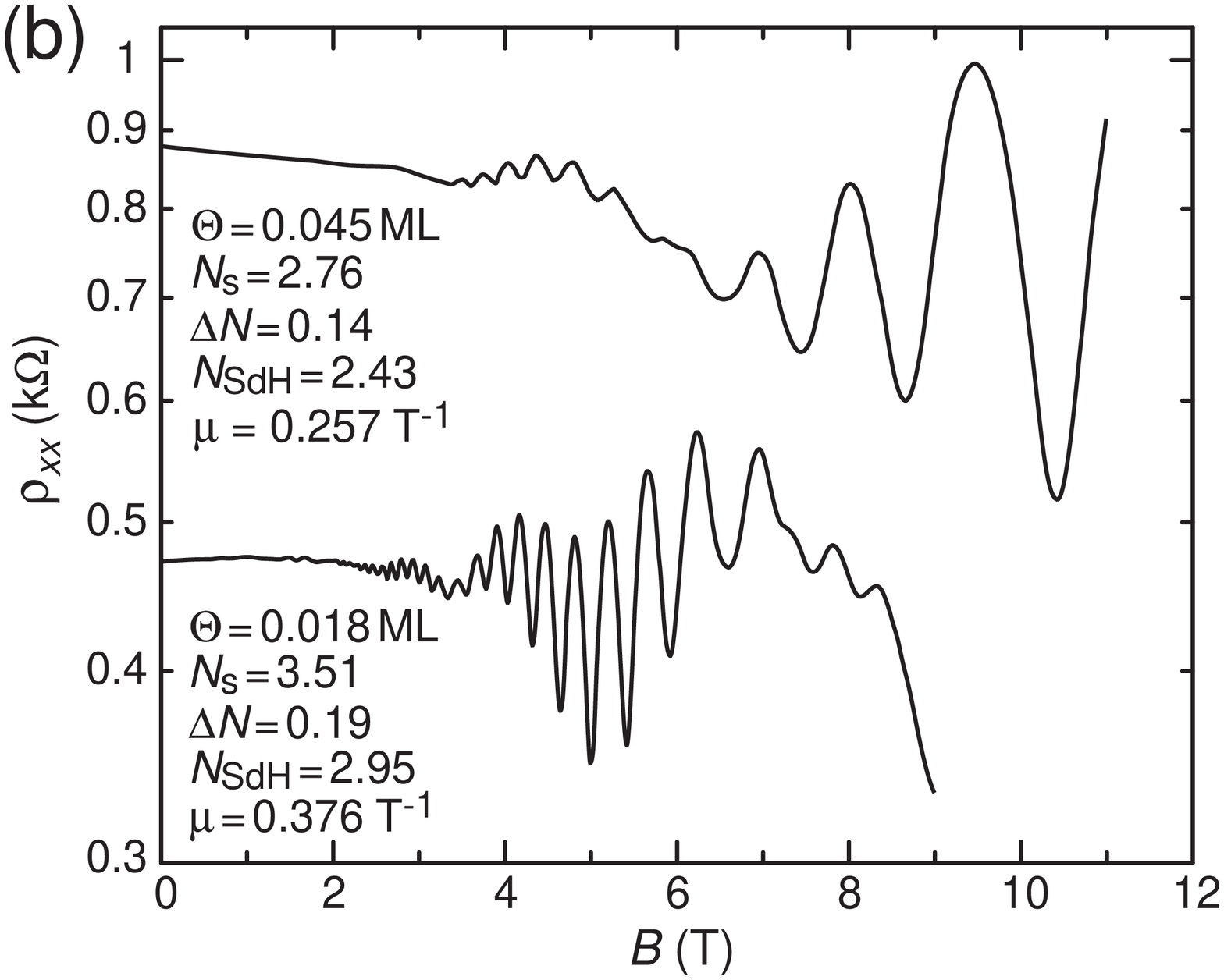}
\caption{
(a) Diagonal resistivity $\rho_{xx}$ (left axis) and Hall resistance $R_{\text H}$ (right axis) for $\Theta=0.18~{\text {ML}}$ with $N_s = 1.04 \times 10^{12}~{\text {cm}}^{-2}$.  $\rho_{xx}$ has minima at $\nu= {\text {even}}$ and the quantum Hall effect is observed for $\nu=4$. (b)  Beating patterns in SdH oscillations for $\Theta=0.018~{\text {ML}}$ with $N_s = 3.51 \times 10^{12}~{\text {cm}}^{-2}$ and $\Theta=0.045~{\text {ML}}$ with $N_s = 2.76 \times 10^{12}~{\text {cm}}^{-2}$.
}
\end{figure}

In the higher $N_s$ regime, the SdH oscillations develop a beating pattern as shown in Fig.~4(b). It is attributed to the Rashba effect \cite{Rashba} based on the spin-orbit interaction in an asymmetric confining potential well. The spin splitting energy is considered to be proportional to the average electric field in the well. This effect has attracted much attention because it essentially concerns a spin-polarized field effect transistor proposed by Datta and Das. \cite{Datta} From the analysis of the SdH oscillations, \cite{Rashba,Dorozhkin} the zero-magnetic-field density difference between spin split two carrier systems was obtained to be $\Delta N = 0.14$ and $0.19 \times 10^{12}~{\rm cm}^{-2}$ for $N_s = 2.76$ and $3.51 \times 10^{12}~{\rm cm}^{-2}$, respectively. The observation of the Rashba spin splitting confirms that the Ag-induced 2D electrons are confined in an inversion layer, of which the average electric field increases with $N_s$. \cite{Ando} According to the calculation for inversion layers, \cite{Merkt} part of electrons are expected to populate the first excited electric subband for $N_s > 1.61 \times 10^{12}~{\rm cm}^{-2}$. Calculated values of the ground electric subband occupation are in good agreement with the electron density deduced from the SdH oscillations $N_{\rm SdH}$ (indicated in Fig.~4(b)), which is lower than $N_s$. 

In conclusion, we have performed Hall measurements on adsorbate-induced electron systems formed at {\it in-situ} cleaved surfaces of $p$-InAs. The Ag-coverage dependence of electron density obtained from the Hall coefficient strongly suggests that each adsorbed Ag atom donates one electron to InAs until the Fermi level at the surface reaches a surface donor level of the adsorbates. The observations of the quantum Hall effect and the Rashba spin splitting demonstrate the perfect two-dimensionality of the system and the strong asymmetry of the confining potential well, respectively. We believe that 2DESs formed at InAs surfaces will provide a fertile ground for the observation of novel transport phenomena. Research on spin-related transport in 2DESs with magnetic adsorbates is promising as well as investigations combined with scanning probe microscopy techniques.

We thank N. Kayukawa for his experimental contributions.
This work is supported by a Grant-in-Aid for Scientific Research
from Japan Society of the Promotion of Science.

\end{document}